# Impact of Wind Generation on Risk-based Security Assessment of Power System

**Umair Shahzad**

School of Computer Science and Engineering,

University of Sunderland, United Kingdom of Great Britain and Northern Ireland

Email: umair.shahzad@sunderland.ac.uk

*Abstract*—The electric power system is one of the largest and most intricate infrastructures. Therefore, it is critical to assess and maintain its security. A power system security assessment is indispensable for identifying post-contingency issues, taking corrective measures, and protecting the system from blackouts. This paper examined the impact of wind generation on the risk-based security assessment of a power transmission network in the context of planning. DIgSILENT PowerFactory software was used to conduct the analysis using a combination of the brute force technique and the nonsequential Monte Carlo (MC) simulation method on the IEEE 39-bus transmission test system. Optimal power flow (OPF) was used to quantify security, considering *(N-1)*, *(N-2)*, and (*N-3)* line outages and an *(N-1)* bus outage. Moreover, the average cost deviation from the mean optimal system operating cost was proposed as a novel security indicator. The results obtianed accurately depicted the effects of changing wind generation levels on system security in terms of risk. The most and least critical line(s) and bus in the system, for different wind generation levels, were also determined. Moreover, the worst-case wind-generation threshold level using two different cost functions for wind was identified.

## NOMENCLATURE

| | |
|---|---|
| $C_i$ | Operating cost (\$/hr) for $i^{th}$ sample of Monte Carlo simulation, without any contingencies, where $i$=1, …, $N_s$. |
| $N_b$ | Number of single bus outages. |
| $\mu_w$ | Mean of wind active power. |
| $C_o$ | Mean optimal system operating cost ($/hr), without any contingencies. |
| $C_\sigma$ | Standard deviation of $C_i$. |
| $C_{\text{cov}}$ | Coefficient of variation of $C_i$. |
| $C_{err}$ | Error of $C_i$. |
| $N_s$ | Number of MC samples. |
| *LF* | Linear wind cost function. |
| *QF* | Quadratic wind cost function. |
| *(L-i)* outages | (*N-i*) line outages, where *i*=1, 2, 3. |
| $C_n^{(L-i)}$ | Mean optimal system operating cost ($/hr) for *(L-i*) outages, where *i*=1, 2, 3; *n*=1, …, $N_i$. |
| $C_n^B$ | Mean optimal system operating cost ($/hr) for single bus outage, where *n*=1, …, $N_i$. |
| $P_r(L_i)$ | Probability of *(L-i)* outage, where *i*=1, 2, 3. |
| $P_r(B)$ | Probability of single bus outage. |
| $R_n^{(L-i)}$ | Risk for *(L-i*) outages, where *i*=1, 2, 3; *n*=1, …, $N_i$. |
| $R_n^B$ | Risk for single bus outage, *n*=1, …, $N_b$. |
| $R_T^{(L-i)}$ | Total risk due to (*L-i*) outages, where *i*=1, 2, 3 *(LF).* |
| $R_T^B$ | Total risk due to single bus outage *(LF).* |
| $\Delta C_{AVG}^B$ | Average cost deviation due to single bus outage *(LF).* |
| $\Delta C_{AVG}^{(L-i)}$ | Average cost deviation due to *(L-i)* outages, where *i*=1, 2 ,3 *(LF).* |
| $R_{TQ}^{(L-i)}$ | Total risk due to (*L-i*) outages, where *i*=1, 2 ,3 *(QF).* |
| $R_{TQ}^B$ | Total risk due to single bus outage *(QF).* |

$\Delta C_{AVGQ}^{(L-i)}$ Average cost deviation due to *(L-i)* outages, where *i*=1, 2 ,3 (*QF)*.

$\Delta C_{AVGQ}^{B}$ Average cost deviation due to single bus outage *(QF)*.

$N_i$ Number of *(L-i)* outages, where *i*=1, 2, 3.

$\sigma_W$ Standard deviation of wind active power.

$\mu_L$ Mean of load active power.

$\sigma_L$ Standard deviation of load active power.

## I. INTRODUCTION

Reliability evaluation of bulk electric power systems (BEPS) can be categorized into two aspects: adequacy and security. BEPS adequacy involves the ability to ensure that there are enough generation facilities in the system to satisfy customer load demands while considering scheduled and rationally anticipated unscheduled outages of system components [1]. Probabilistic methods have been used for traditional BEPS adequacy assessment [2-3]. Security is defined as the ability of the system to endure sudden disturbances, such as electrical short circuits or the unexpected loss of system components. It also considers system operating conditions and the probability of disturbances. BEPS security assessment usually deals with operation of the system in various states: normal, alert, emergency, and extreme emergency [4].

Power systems must operate according to some constraints for secure operation. These constraints are called equality and inequality constraints. The equality constraint states that the total power generated must be equal to demand plus the system losses. The inequality constraint includes generator active and reactive power limits, transmission line mega-volt-amperes (MVA) limits, and bus voltage limits. In the normal operating state, the system is said to be secure; and all equality and inequality constraints are satisfied. In the alert state, all equality and inequality constraints are fulfilled; but there is no reserve power generation [4]. Consequently, at least one inequality constraint is not satisfied in the event of a contingency. At this stage, preventive control actions, such as active and reactive power control of generators, must be taken to restore the system to its normal state. In the emergency state, all equality constraints are fulfilled; but at least one inequality constraint is not satisfied. At this point, corrective actions, such as load shedding, tripping of transmission lines, or disconnection of generators, must be done. In an extreme emergency state, the power network switches into an islanded operation mode, where both equality and inequality constraints are not fulfilled. If the system enters this state, it cannot return to the emergency state [4].

Security assessment, in general, uses the traditional deterministic criterion called the *(N-1)* security criterion, in which loss of any single component (known as a contingency), will not cause system failure [5]. Security assessment is further divided into two types: static (steady state) and dynamic (transient) [6]. The goal of the static security assessment is to determine whether, after the manifestation of a disturbance, there is a new steady-state operating point. On the other hand, dynamic security assessment deals with the ability of the power system to reach a stable point when subjected to a severe transient disturbance, such as a three-phase fault on a transmission line, sudden loss of generators, or loss of a large load [7]. The focus of this work is on static security assessment.

Based on the type of analysis, power system security assessment can be grouped into two types: deterministic and probabilistic. In the deterministic method, the analysis is based on the credible and most severe disturbances. The main flaw of using the deterministic approach is that it considers all security issues with an equal risk, which results in a higher cost of operation. In contrast, the probabilistic technique takes the risk into account by considering both the probability and consequence of the contingencies [6-7].

The product of probability of an unforeseen event and its impact is commonly known as risk, which is mathematically defined as (1) [8].

$$Risk(X_{t,f}) = \sum_{i} \mathrm{P_r}(E_i)\Big(\sum_{j} \mathrm{P_r}(X_{t,j} \mid X_{t,f}) \times Sev(E_i, X_{t,j})\Big) \tag{1}$$

where $X_{t,f}$ is the forecasted operating condition at time $t$; $X_{t,j}$ is the $j^{th}$ possible loading level; $\mathrm{P_r}(X_{t,j} \mid X_{t,f})$ is the probability of $X_{t,j}$ given $X_{t,f}$ has occurred, which is obtained using a probability distribution for the possible load levels; $E_i$ is the $i^{th}$ contingency, and $\mathrm{P_r}(E_i)$ is its probability; and $Sev(E_i, X_{t,j})$ quantifies the impact of $E_i$ occurring under the $j^{th}$ possible operating condition.

The current industry practices use the deterministic approach for security assessment. Although the deterministic approaches result in highly secured power systems, they do not consider the probability of operating conditions. Apart from the high cost due to conservative designs, the key drawback with the deterministic assessment techniques is that they consider all security problems to have equal risk. Moreover, the integration of renewable generation introduces more stochasticity in the system, making the application of probabilistic practices essential in the security assessment process. Probabilistic assessment methods can be very effective when uncertain parameters are the prime characteristics of the power system. It is, thus, of great implication to propose a risk-based security assessment approach, to overcome the inadequacies of the deterministic approach. With the evolution of renewable energy sources, such as wind and solar energy, and the demand for sustainable energy, probabilistic risk assessment is gaining popularity [9]. The main motivation of using a risk-based approach is that it merges the probability of occurrence of an unforeseen event with the impact of the event, thereby providing a profound understanding of the security of the power system [10]. The higher the risk, the lower the system security and vice versa. Thus, the key motivation of the proposed approach in this paper is to augment the current deterministic security assessment practices to cater for the future needs of the power systems (which will include abundant uncertainty and renewable generation).

The chief objective of this research paper is to analyze the impact of wind generation penetration on static security of a transmission network, in terms of risk, using a combination of the brute force method, and the nonsequential Monte Carlo (MC) simulation method. The analysis incorporates a single bus outage in addition to *(L-i)* outages, as contingency events. The average cost deviation from the mean optimal operating cost of the system is proposed as a system static security indicator; and subsequently, this is used to quantify the impact for computing risk.

To the best of author's knowledge, none of the research papers, on transmission system security assessment, has quantified security in terms of economics. Majority of existing works on probabilistic static security assessment ignored bus outages and considered only single line outages. Also, the impact of high wind penetration was overlooked in many works. Hence, in this paper, average deviation from mean optimal operating cost of the system is considered as a system security indicator and subsequently, this is used to quantify the impact for computing risk. Moreover, buses and transmission lines are ranked in terms of risk. The key novelty of the work is the quantification of system security terms of a novel indicator, i.e., average deviation from mean optimal operating cost and consequently, determining the risk for security analysis, considering multiple outages [*(N-i)* line outages (where $i>1$), and *(N-1)* bus outages]. MC simulation approach is used for sampling the probability density functions (PDFs) of system loads and wind generation.

## II. Optimal Power Flow Formulation Incorporating Wind Generation

Optimal power flow (OPF) is an optimization tool for power system operation analysis. The OPF procedure attempts to optimize an objective function, ensuring that available generation is allocated to satisfy the load demand while keeping all power system components within their specified limits to satisfy the equality constraints (total power loss equals total generation minus total demand) and inequality constraints (i.e., generator active and reactive power limits, branch MVA limits, and bus voltage limits). The most common OPF objective function is the minimization of operating cost. In this paper, minimization of mean system operating cost is used as the objective function. However, if there is infeasibility, i.e., the OPF does not converge, minimization of load shedding is used instead. If the optimization is done by minimizing load shedding, there is a huge penalty associated with it.

The cost function of thermal generators is usually in the form of a quadratic equation [11]. In this paper, it is assumed that the independent system operator (ISO) is buying wind energy from an independent power producer (IPP) [12]. Therefore, the wind cost is incorporated as an operating cost in the system. A linear cost function, based on parameters from [13], is used to represent the wind cost. Moreover, the analysis is done using the quadratic wind cost function for comparative purposes and verification of results obtained with the linear wind cost function. The assumption for using quadratic cost functions for wind generation is in accordance with [14]. It must, however, be noted that the wind generation cost function depends on the IPP; and it can be in any form based on the mutual understanding of the buyer (ISO) and seller (IPP).

A normal (Gaussian) distribution is used to represent the randomness of wind active power as suggested by [15]. The normal probability density function (PDF) is a good model when applied to quantities that are likely to be the sum of several independent situations. Thus, this is a reasonable assumption as the output power of a wind generator is influenced by various independent events, such as air density, temperature, pressure, and humidity [15].

## III. Security-Constrained Optimal Power Flow: Overview and Previous works

Security-constrained OPF (SCOPF), in general, deals with OPF in the presence of *(N-k)* a power system elements outage, where *k* denotes the number of elements which are out of service. The secure operation of a power system requires that there are no uncontrollable contingency violations. Thus, in this case, the minimization of the objective function is done considering contingencies. The SCOPF adjusts the controls to the base case (precontingency condition) to avert violations in the post-contingency conditions [16]. [17] proposed a methodology to solve the preventive risk based SCOPF problems. A two-stage Benders decomposition strategy was proposed in this regard. [18] proposed a heuristic approach to compute the worst-case under operation uncertainty for a contingency with respect to overloads. The problem was formulated as a non-convex nonlinear bilevel program. [19] proposed an enhanced risk-based AC SCOPF problem formulation to aid transmission system operators (TSOs) trading-off economy and (thermal) security. The system security was controlled through an overall system risk constraint.

In [20], a risk model was presented for risk related to outages, accounting for available remedial measures and the impact of cascading events. [21] proposed a risk-based contingency-constrained OPF model by leveraging the methods of both adjustable uncertainty set and distributionally robust optimization. In the proposed model, an adjustable uncertainty set of wind power was established with network contingencies explicitly incorporated. [22] suggested risk based static security assessment on a practical interconnected power system at various loading condition using a risk index. The result captured the uncertainty in the occurrence of contingency and the variation of risk index as the load changes. [23] described a mathematical framework for the solution of the economic dispatch problem with security constraints, which considered the system corrective capabilities after the outage occurred. [16] presented an approach for OPF with incorporation of preventive and corrective control. The approach is useful for operation planning to guarantee secure operation under both normal and contingency states. [24] proposed a new optimization tool based on the cross-entropy method to assess SCOPF solutions. Different types of density functions were tested to cope with discrete variables present in the SCOPF problem. Besides the works mentioned above, an avid reader can refer to [25-26] for detailed background about research works associated with SCOPF.

The rest of the paper is organized as follows. Section IV describes the computation procedure for risk-based security assessment. Section V outlines the mathematical formulation for the proposed security indicator. Sections VI and VII discuss case studies and simulations and the associated results, respectively. Section VIII concludes the paper with proposed future research directions.

## IV. Computation Procedure for Risk-Based Security Assessment

In this paper, the average cost deviation from the mean optimal system operating cost is proposed as an indicator for assessing system security. The main goal is to examine the impact of wind generation penetration levels on system security, in terms of risk, and to determine the worst-case wind generation threshold level (based on maximum average cost deviation and maximum risk). To assess security, SCOPF is conducted, considering *(L-i)* and single bus outages; and the resulting average cost deviations are determined. Consequently, system security is assessed based on average cost deviations. The computation procedure for risk-based security assessment is outlined in Fig. 1.

There are five different cases considered. First, the analysis is conducted assuming linear wind cost parameters. The following discussion, regarding steps, applies to all five cases. In the first step, OPF is conducted without considering any line or bus contingencies. The MC method is used to sample the PDFs of loads and wind generation. In the second and third steps, OPF is performed considering *(L-i)* and single bus outages using the brute force method. Using $C_o$, average cost deviations and, consequently, risk values for *(L-i)* and single bus outages are computed. Consequently, the least and most critical line(s) and the least and most critical bus are determined. In the next step, the impact of wind generation levels on the risk values of *(L-i)* and single bus outages is observed. Moreover, the impact on the average cost deviation of the changing wind generation penetration is observed. The worst-case wind generation threshold level (this level indicates the maximum percentage of conventional generation that can be replaced by wind generation) is then determined based on the analysis conducted. Finally, the entire process of determining the worst-case wind generation threshold level is repeated using a quadratic wind cost function.

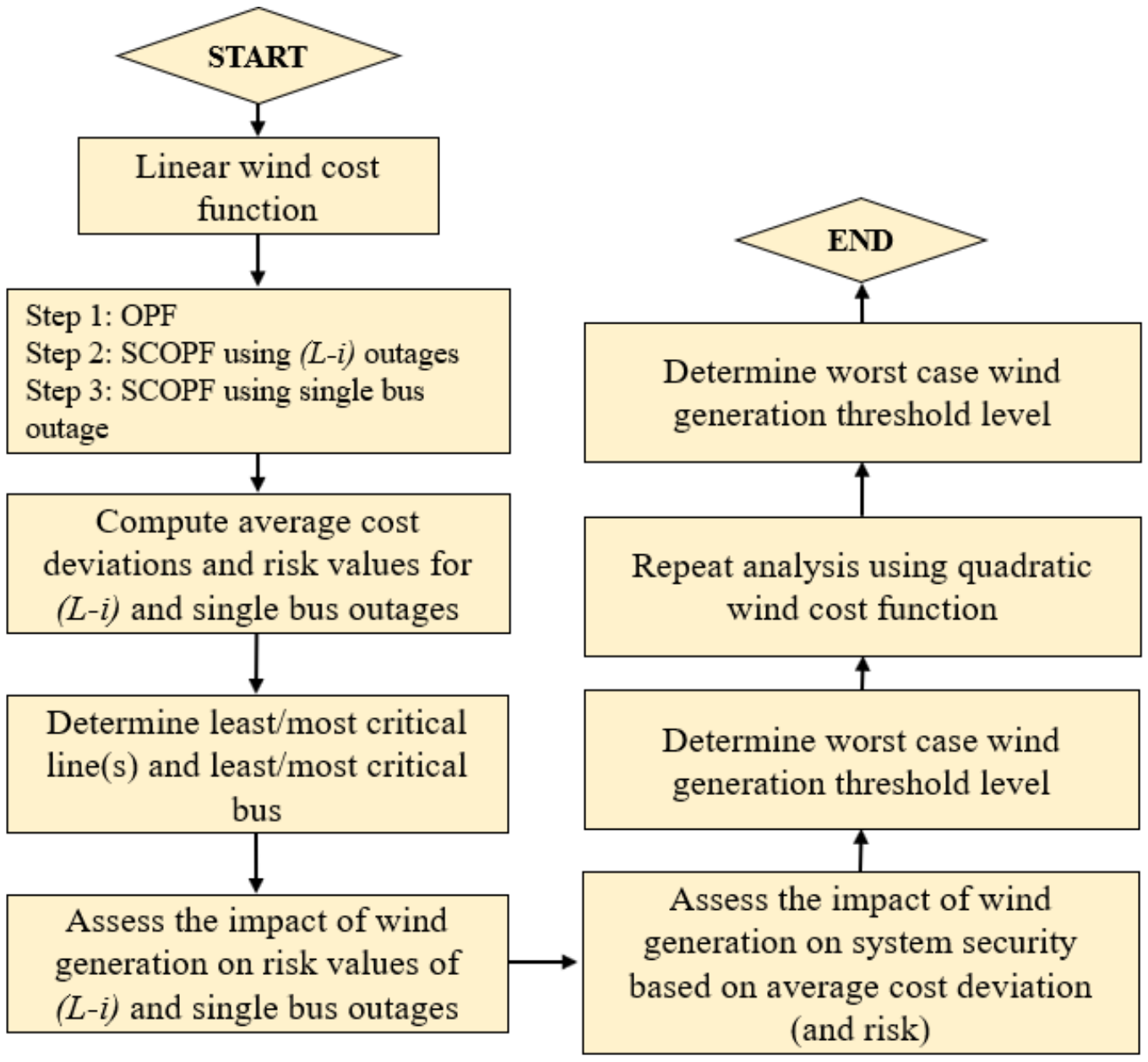

Fig. 1. Computation procedure for risk-based static security assessment

## V. MATHEMATICAL FORMULATION FOR THE PROPOSED SECURITY INDICATOR

This section describes the mathematical formulation for the proposed security indicator. First, $R_n^{(L-i)}$ is computed by:

$$R_n^{(L-i)} = \mathrm{P_r}(L_i) \times \Delta C_{OL}^{\,n} \tag{2}$$

where $\Delta C_{OL}^{\,n}$ is defined as:

$$\Delta C_{OL}^{\,n} = C_n^{(L-i)} - C_o \tag{3}$$

It is assumed that the values of $\mathrm{P_r}(L_i)$ are obtained considering historical data and maintenance practices.

$R_T^{(L-i)}$ is computed as:

$$R_T^{(L-i)} = \sum_{n=1}^{N_i} R_n^{(L-i)} = \sum_{n=1}^{N_i} \mathrm{P_r}(L_i) \times \Delta C_{OL}^{\,n} \tag{4}$$

$R_n^{B}$ is calculated as:

$$R_n^{B} = \mathrm{P_r}(B) \times \Delta C_{OB}^{\,n} \tag{5}$$

where $\Delta C_{OB}^{\,n}$ is defined by:

$$\Delta C_{OB}^{\,n} = C_n^{B} - C_o \tag{6}$$

It is assumed that the value of $\mathrm{P_r}(B)$ is obtained considering historical data and maintenance practices.

The total risk due to a single bus outage, $R_T{}^B$, is computed as:

$$R_T{}^B = \sum_{n=1}^{Nb} R_n{}^B = \sum_{n=1}^{Nb} \mathrm{P_r}(B) \times \Delta C_{OB}{}^n \quad (7)$$

The average cost deviation due to *(L-i)* outages, $\Delta C_{AVG}{}^{(L-i)}$, is calculated as:

$$\Delta C_{AVG}{}^{(L-i)} = \sum_{n=1}^{N_i} \frac{\Delta C_{OL}{}^n}{N_i} \quad (8)$$

$\Delta C_{AVG}{}^B$ is computed as:

$$\Delta C_{AVG}{}^B = \sum_{n=1}^{N_b} \frac{\Delta C_{OB}{}^n}{N_b} \quad (9)$$

$\Delta C_{AVG}{}^{(L-i)}$ represents an indicator for security, considering *(L-i)* outages. Similarly, $\Delta C_{AVG}{}^B$ is an indicator for security, considering a single bus outage. The higher these values, the lower the system security and vice versa.

## VI. CASE STUDIES AND SIMULATIONS

The IEEE 39-bus test transmission system, operating at 345 kV, was used to conduct the required simulations. Its single line diagram is shown in Fig. 2. The system has 34 transmission lines. The numerical data and parameters were taken from [27]. A normal distribution is used to represent the randomness of wind active power. The $\mu_W$ was chosen to be the original value of thermal generation active power, which is replaced, and $\sigma_W$ was chosen to be 10% of the mean value. Similarly, a normal distribution is used to define the uncertainty in system loads. The active power of each load was assigned a $\mu_L$ equal to the original load active power value, as given in test system data in [27], and a $\sigma_L$ equal to 10% of the mean value. The coefficients of generator cost curves for thermal and wind generators were taken from [28] and [13], respectively. For this research, the load shedding cost penalty was assumed to be $10,000 per MW of lost load.

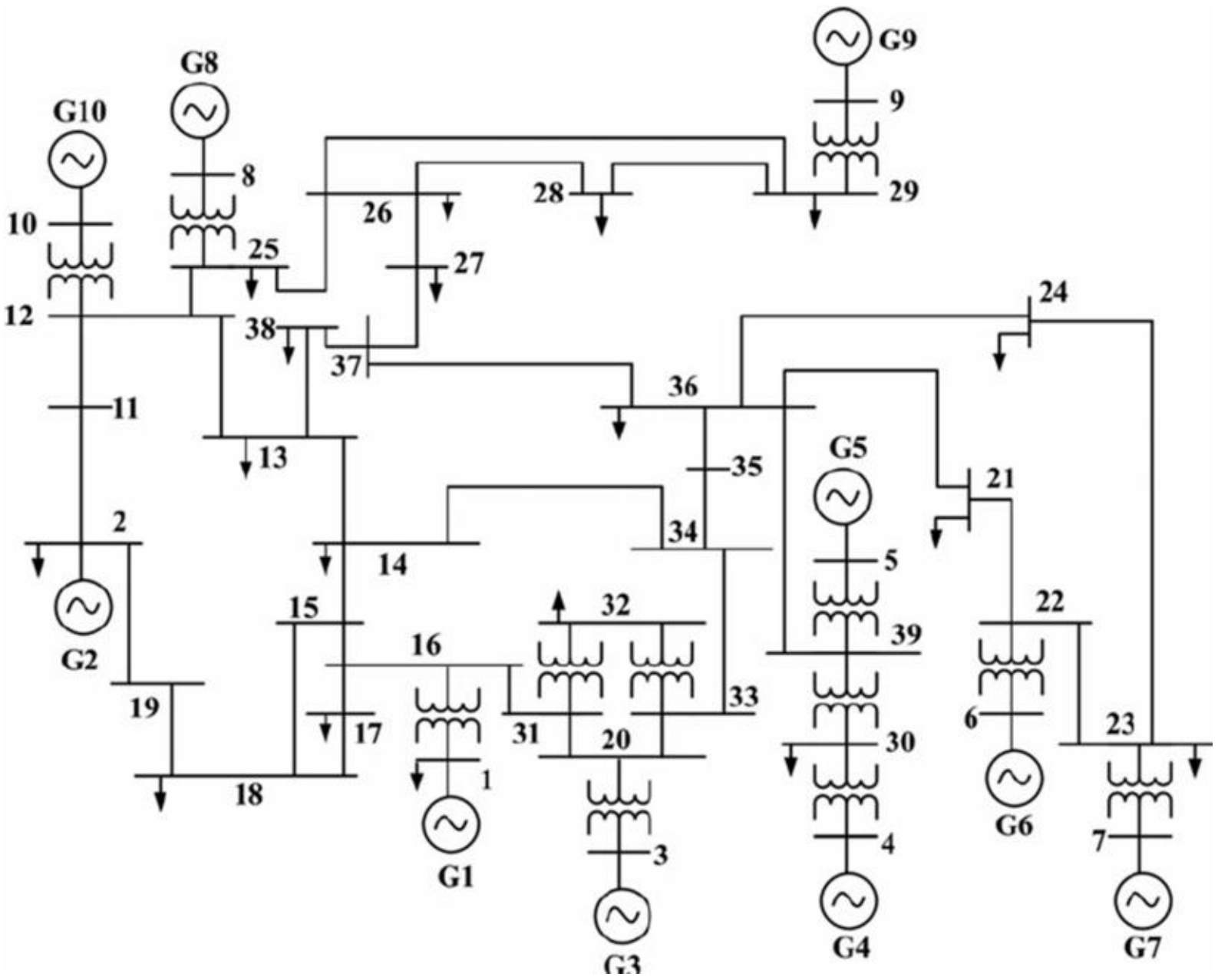

Fig. 2. IEEE 39-bus test system

The description of the five cases considered is as follows. In Case 1, the test system was used in its original format, i.e., no wind generation was present in the network and all generation consisted of conventional thermal synchronous generators. Cases 2-5 considered 26%, 52%, 82%, and 100% wind penetration, respectively. This percentage indicates that this amount of wind generation was used to replace the original thermal generation in the network (for instance, 26% wind penetration implies that 26% of the original thermal generation active power was replaced by equivalent wind generation, and similarly for other cases). Moreover, it was assumed that the area was rich in wind resources wherever wind was connected in the network. Also, it was assumed that wind generators are equipped with sufficient voltage and frequency controls to keep the system stable. As we are dealing with static security and evaluation of OPF following contingencies, the dynamics of wind generators were ignored. The value of $\Pr(L_i)$ was assumed to be $10^{-2}$, $10^{-4}$, and $10^{-6}$, for $i$=1, 2, and 3, respectively. The value of $\Pr(B)$ was assumed to be $10^{-7}$. Using the brute force method, the total number of *(L-i)* outages considered was 34, 561, and 5984, for $i$= 1, 2, and 3, respectively; and the total number of single bus outages considered was 39. DIgSILENT PowerFactory software was used to conduct the required simulations [29].

## VII. Results and Discussion

First, the number of MC samples, $N_s$, needed to be determined. Various MC simulations with different values of $N_s$ were conducted on Case 1. The results obtained are shown in Table I. As is evident, after $N_s$=1000, $C_o$, $C_\sigma$, and $C_{\text{cov}}$ did not change. Thus, $N_S$=1000 was selected for the case studies. Following the computation procedure described in Fig. 1, OPF and SCOPF results were obtained for Cases 1-5. The values of $C_o$, $C_\sigma$, $C_{cov}$, and $C_{err}$ for each case, are shown in Table II.

TABLE I. DETERMINATION OF $N_S$ FOR MC SIMULATIONS

| $N_s$ | $C_o$ | $C_\sigma$ | $C_{cov}$ | $C_{err}$ |
|---|---|---|---|---|
| 100 | 4,465,004 | 143,331 | 0.03 | 14,333 |
| **1,000** | **3,125,486** | **94,193** | **0.03** | **2,978** |
| 10,000 | 3,125,486 | 94,193 | 0.03 | 941 |

TABLE II. VALUE OF $C_O$ FOR DIFFERENT CASES ($N_S$=1000)

| **Case No.** | **Case Type** | $C_o$ | $C_\sigma$ | $C_{cov}$ | $C_{err}$ |
|---|---|---|---|---|---|
| 1 | No Wind | 3,125,486 | 94,193 | 0.03 | 2,978 |
| 2 | 26% Wind | 2,232,163 | 67,405 | 0.03 | 2,131 |
| 3 | 52% Wind | 398,602 | 12,286 | 0.03 | 388 |
| 4 | 82% Wind | 152,047 | 4,568 | 0.03 | 144 |
| 5 | 100% Wind | 23,747 | 717 | 0.03 | 23 |

For each case, OPF was conducted considering *(L-i)* and single bus outages. Consequently, values of $R_n^{(L-i)}$ and $R_n^B$ were determined. As an illustration, Fig. 3 and 4 show the graphical variation of $R_n^{(L-1)}$ and $R_n^B$, respectively, with different wind generation penetration levels. Similar graphs can be obtained for $R_n^{(L-2)}$ and $R_n^{(L-3)}$. In Fig. 3 and 4, L1-2 indicates that the line is connected between Buses 1 and 2; similarly, B1, B2, etc., indicate bus numbers. These graphs are helpful to study the impact of wind penetration levels on the failure of a specific bus or line, in terms of risk. For all the cases, the most and least critical line(s) and most and least critical buses, on average, were found, based on $R_n^{(L-i)}$ and $R_n^B$ values. The most and least critical line(s) are the ones with maximum and minimum values of $R_n^{(L-i)}$, respectively. Similarly, the least and most critical buses are the ones with minimum and maximum $R_n^B$ values, respectively. For instance, referring to Fig. 3, in Case 3, Line L8-9 is the most critical; and Line L7-8 is the least critical. Similarly, referring to Fig. 4, Bus 39 and Bus 37 are the most and least critical buses, respectively (Case 3).

The results, including those from the analysis of *(L-2)* and *(L-3)* outages, are summarized in Tables III and IV. These results are vital for a power system planner as they give an idea of which transmission line(s) or buses to give more significance during decision making. From Table IV, it is evident that Bus 39 is always the most critical bus for any wind penetration level. It must be noted that the quantification of criticality highly depends on the values of $\mathrm{P_r}(L_i)$ and $\mathrm{P_r}(B)$, cost parameters of synchronous/wind generators, and load demand. Therefore, for a practical test system, these values must be estimated accurately. Moreover, it is also interesting to study the values $R_T^{(L-i)}$ and $R_T^B$ for all cases. Fig. 5 shows the variation of these values with wind generation levels. As evident from Table II, $C_o$ decreased as wind penetration increased on the network because the cost of wind generation was assumed to be less than that of conventional thermal generators.

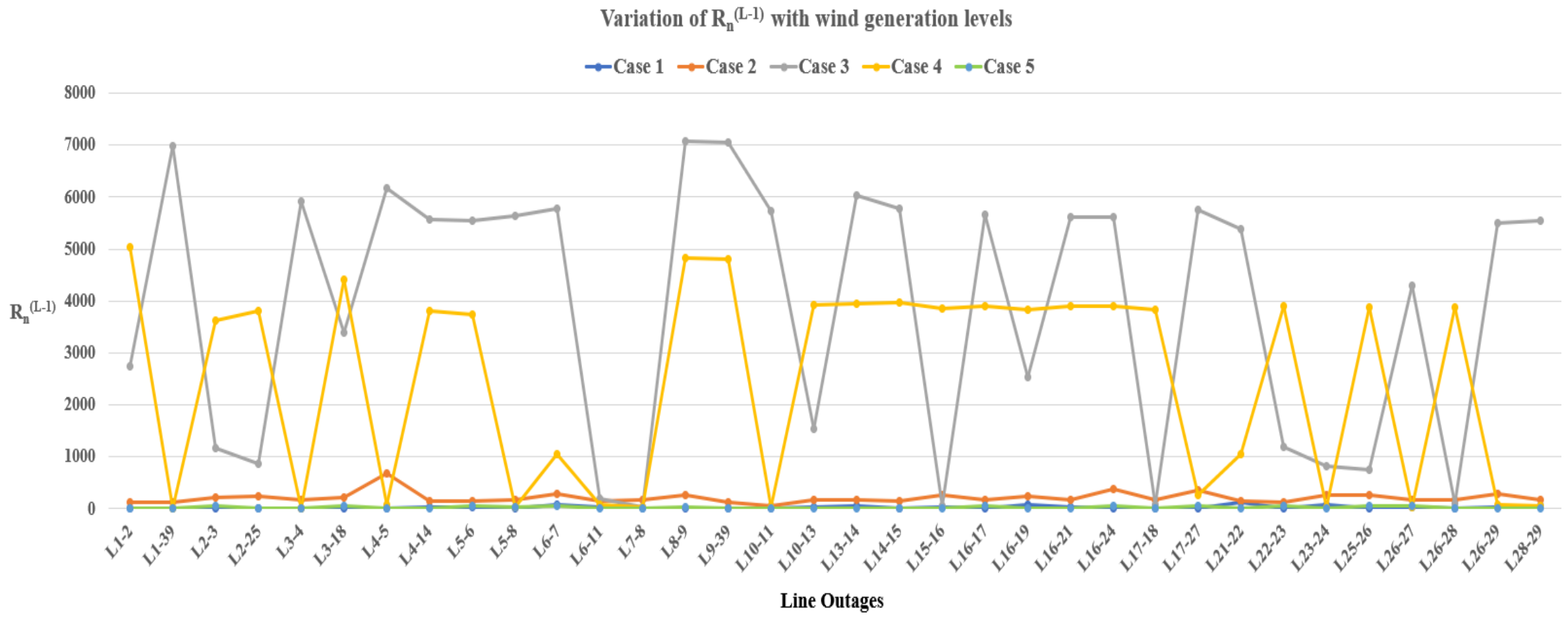

Fig. 3. Variation of $R_n^{(L-1)}$ with wind generation levels

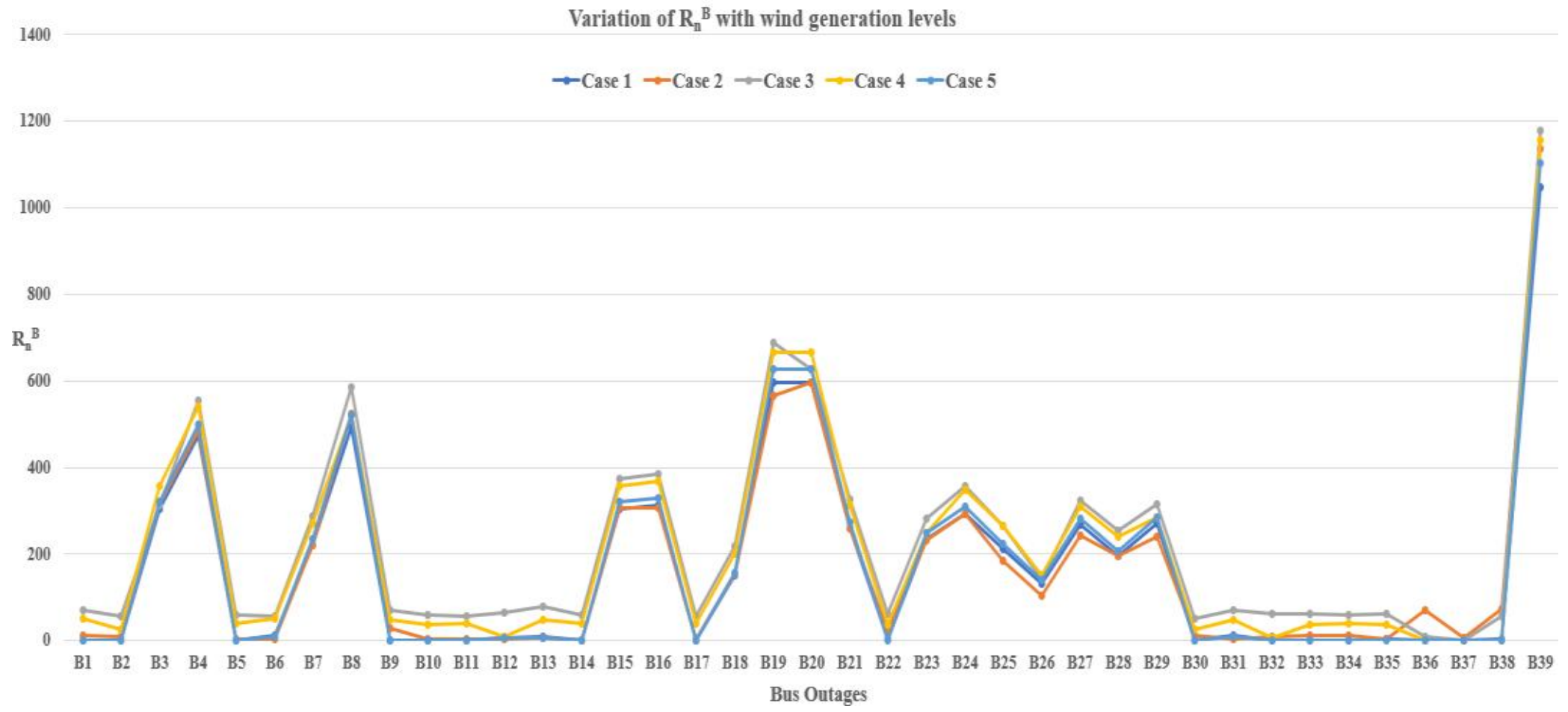

Fig. 4. Variation of $R_n{}^B$ with wind generation levels

Referring to Fig. 5, when there is no wind, $R_T{}^B$ was greater than $R_T{}^{(L-i)}$. However, with the increase in wind, $R_T{}^B$ became less than $R_T{}^{(L-i)}$.This trend continued until 82% wind penetration. However, at 100% wind, $R_T{}^B$ again became greater than $R_T{}^{(L-i)}$. The values of $R_T{}^{(L-i)}$ and $R_T{}^B$ increased until 52% wind penetration; however, after that they decreased significantly. A possible reason is that there was more wind in the network which resulted in lower cost; and hence, the risk values decreased.

TABLE III. RANKING OF LINES BASED ON $R_n^{(L-1)}$ VALUES

| Case No. | Most Critical (L-1) Line | Least Critical (L-1) Line | Most Critical (L-2) Lines | Least Critical (L-2) Lines | Most Critical (L-3) Lines | Least Critical (L-3) Lines |
|---|---|---|---|---|---|---|
| 1 | L21-22 | L25-26 | L14-15 | L16-24 | L7-8 | L13-14 |
| | | | L17-18 | L26-27 | L10-13 | L26-28 |
| | | | | | L21-22 | L28-29 |
| 2 | L4-5 | L10-11 | L4-14 | L13-14 | L4-14 | L10-11 |
| | | | L9-39 | L17-18 | L14-15 | L14-15 |
| | | | | | L17-18 | L16-24 |
| 3 | L8-9 | L7-8 | L9-39 | L8-9 | L9-39 | L8-9 |
| | | | L22-23 | L16-19 | L14-15 | L10-13 |
| | | | | | L28-29 | L15-16 |
| 4 | L1-2 | L3-4 | L2-25 | L3-18 | L2-25 | L3-18 |
| | | | L14-15 | L4-14 | L4-5 | L4-14 |
| | | | | | L16-19 | L6-11 |
| 5 | L6-7 | L16-21 | L6-11 | L16-24 | L6-7 | L16-24 |
| | | | L17-27 | L26-28 | L8-9 | L17-27 |
| | | | | | L16-21 | L22-23 |

TABLE IV. RANKING OF BUSES BASED ON $R_n^B$ VALUES

| Case No. | Most Critical Bus | Least Critical Bus |
|---|---|---|
| 1 | B39 | B1 |
| 2 | B39 | B14 |
| 3 | B39 | B37 |
| 4 | B39 | B36 |
| 5 | B39 | B14 |

In other words, thermal generation dominates the objective function until 52% wind; and after that, wind generation takes over. Moreover, $R_T^{(L-3)}$ is always greater than $R_T^{(L-2)}$ and $R_T^{(L-1)}$ at each wind generation level since *(L-3)* outages have a much more greater impact than *(L-1)* and *(L-2)* outages, i.e., $\Delta C_{AVG}^{(L-3)} > \Delta C_{AVG}^{(L-1)}$, and $\Delta C_{AVG}^{(L-2)}$.

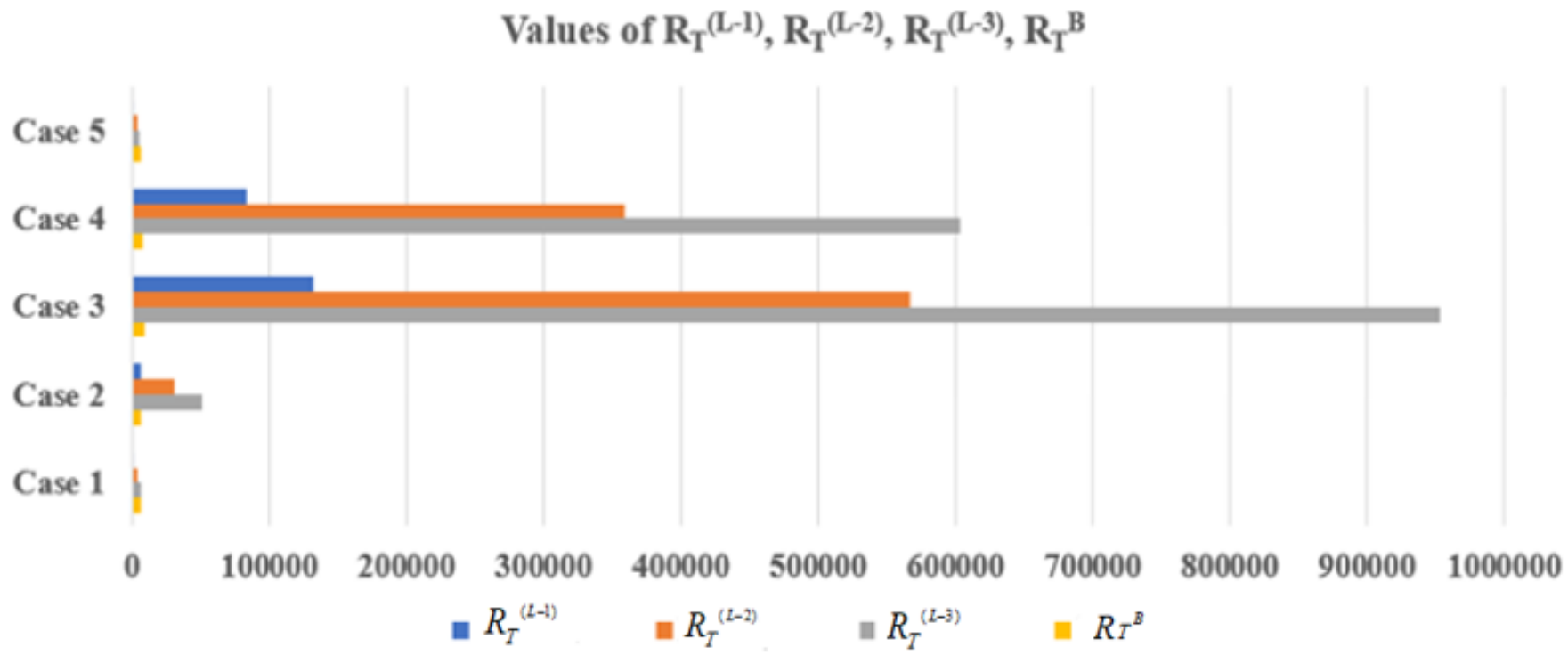

Fig. 5. Values of $R_T^{(L-1)}$, $R_T^{(L-2)}$, $R_T^{(L-3)}$, and $R_T^B$

Analyzing the variation of average cost deviations is also thought provoking. Fig. 6 shows the graphical variation of average cost deviations for *(L-i)* and single bus outages. As evident from Fig. 6, values of $\Delta C_{AVG}^{(L-i)}$ and $\Delta C_{AVG}^{B}$ increased until 52% wind and then decreased drastically. The reason is the same as explained in the case of the trend followed by $R_T^{(L-i)}$ and $R_T^B$ in Fig. 5. Moreover, for each case, the impact $\Delta C_{AVG}^{B}$ was always greater than $\Delta C_{AVG}^{(L-i)}$ and for the same wind penetration level. It is interesting to compare $\Delta C_{AVG}^{(L-3)}$ with $\Delta C_{AVG}^{B}$. As $\Delta C_{AVG}^{B}$ was always greater than $\Delta C_{AVG}^{(L-3)}$, this implies that although $P_r(B) < P_r(L_3)$, the impact of a single bus outage, on average, is much greater than the simultaneous outage of three lines.

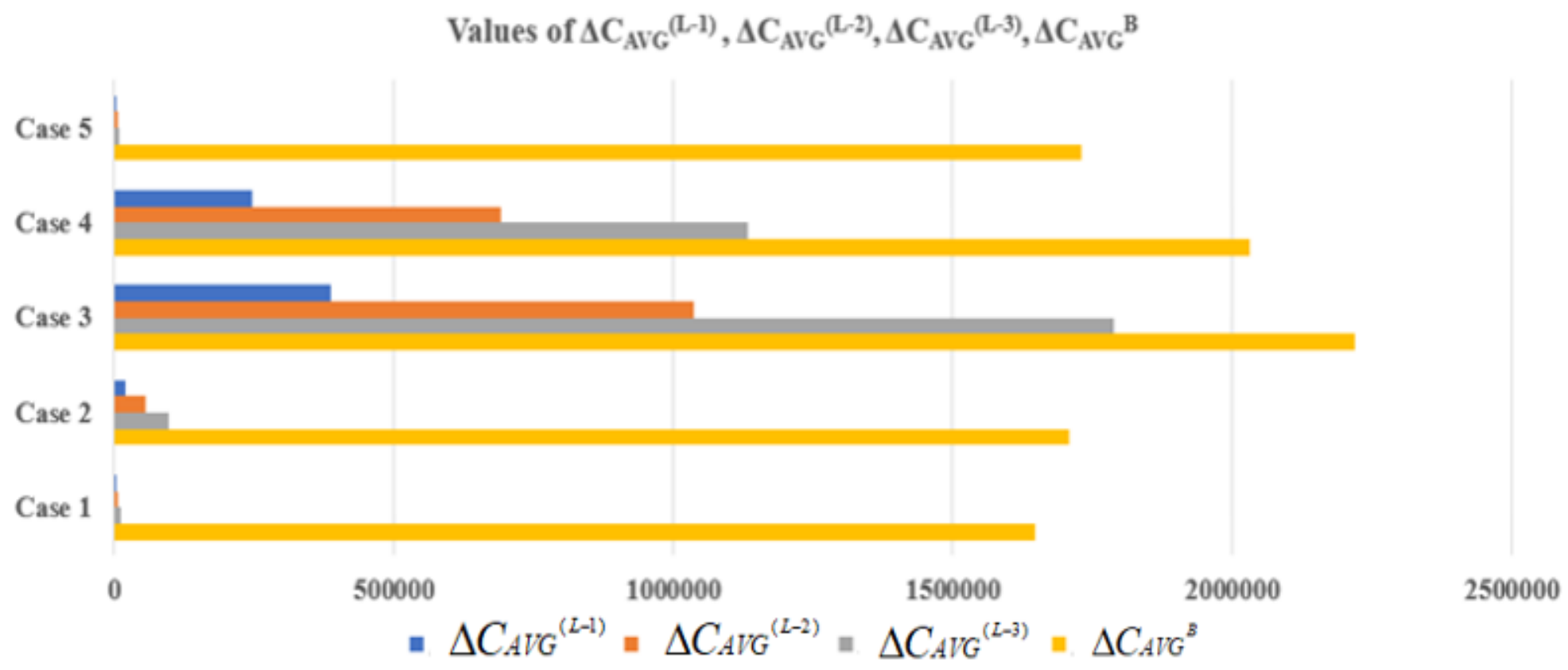

Fig. 6. Values of $\Delta C_{AVG}^{(L-1)}$, $\Delta C_{AVG}^{(L-2)}$, $\Delta C_{AVG}^{(L-3)}$, and $\Delta C_{AVG}^{B}$

From Fig. 5 and Fig. 6, it can be deduced that the 52% wind generation level is the worst-case threshold level, as it resulted in maximum average cost deviation and maximum risk values for *(L-i)* and single bus outages. In the last step, average cost deviations and risk values for *(L-i)* and single bus outages were computed using a quadratic wind cost function. The graphical results are shown in Fig. 7 and 8. As evident, even using a quadratic wind cost, as opposed to linear wind cost function, did not change the trends in results. The worst-case wind threshold level was still 52%. The only difference was that values of $\Delta C_{AVG}^{(L-i)}$, $\Delta C_{AVG}^{B}$, $R_T^{(L-i)}$, and $R_T^{B}$ were less than their corresponding quadratic counterparts. This is understandable as quadratic function has an added cost coefficient when compared to linear function.

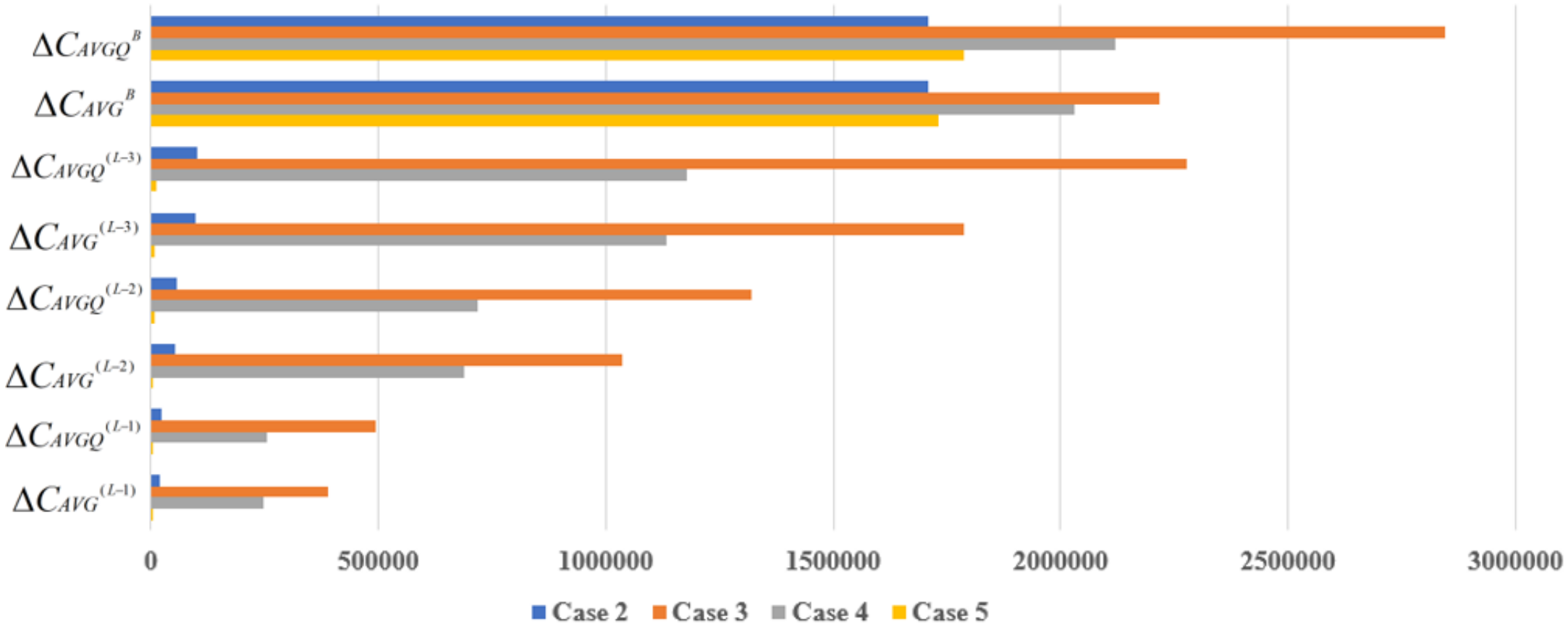

Fig. 7. Comparison of average cost deviations for linear and quadratic wind cost functions

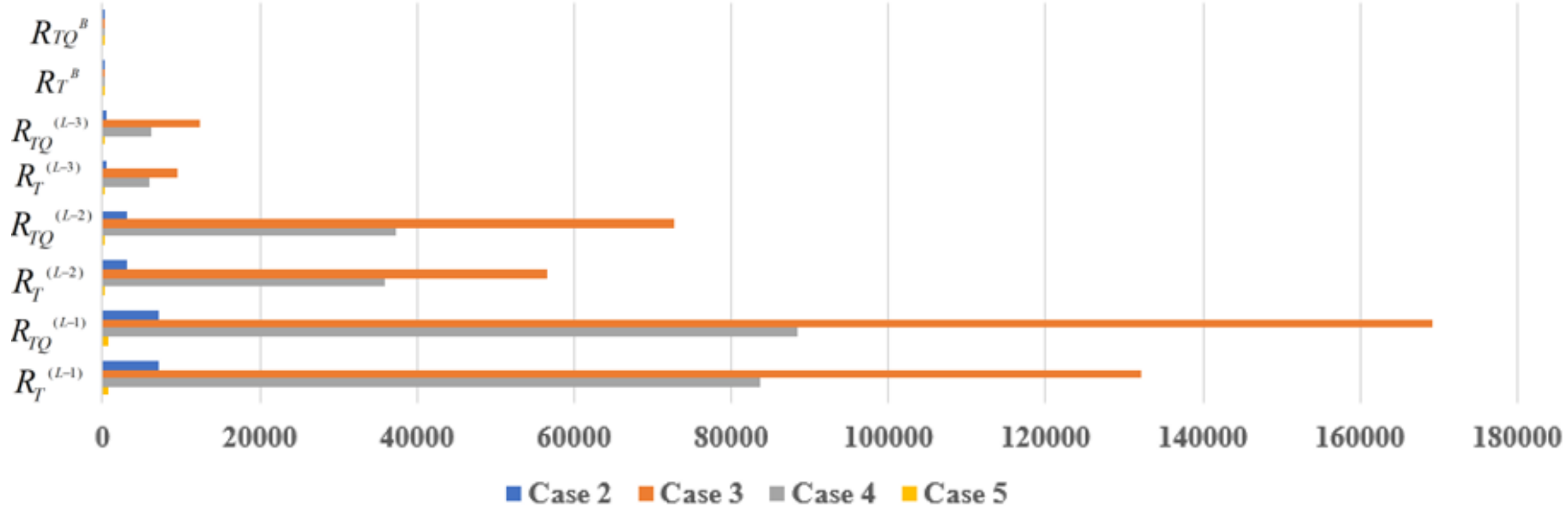

Fig. 8. Comparison of risks for linear and quadratic wind cost functions

The risk values $R_T^{(L-i)}$ and $R_T^{B}$ increased up to 52% wind penetration. However, after that, there was a drastic decrease. This decrease was due to the reduced dominance of thermal synchronous generators after 52% wind penetration. As wind cost is assumed to be less than the cost of thermal synchronous generators, the reduction in risk value occurs. Moreover, values of $\Delta C_{AVG}^{(L-i)}$ and $\Delta C_{AVG}^{B}$ followed the same trend. It was determined that a 52% wind penetration level is the worst-case threshold level (for both linear and quadratic wind cost functions). Thus, if 52% of thermal generation is replaced by wind, it is detrimental to system security. Higher wind penetrations gave productive results, as values of $R_T^{(L-i)}$, $R_T^B$, $\Delta C_{AVG}^{(L-i)}$, and $\Delta C_{AVG}^B$ decreased. The same trend was observed if comparatively lower (26%) wind penetration was used. However, values of $R_T^{(L-i)}$, $R_T^B$, $\Delta C_{AVG}^{(L-i)}$, and $\Delta C_{AVG}^B$ for 100% wind were in close proximity to that of the base case.

For Cases 1 and 5, $R_T^B$ was greater than $R_T^{(L-i)}$, but $\Delta C_{AVG}^B$ was always greater than $\Delta C_{AVG}^{(L-i)}$ for all cases. This implies that the impact of a single bus outage, on average, was much greater than those of line outages (1, 2, or 3 at a time). Moreover, for Cases 2, 3, and 4, total risk due to a single bus outage was less than the total risk due to (*L-i*) outages (i.e., $R_T^B < R_T^{(L-i)}$) but their impact $\Delta C_{AVG}^B$ was greater than $\Delta C_{AVG}^{(L-i)}$.

Therefore, in addition to *(L-i)* outages, it is important to consider a single bus outage in power system static security assessment. For different wind penetration levels, most (and least) critical line(s) and the most (and least) critical buses were identified based on $R_n^{(L-i)}$ and $R_n^B$ values. This is important, as it can provide a system planner with a quick idea of which lines/bus should be given the most importance during decision-making. Finally, graphical variation of values of $R_n^{(L-1)}$ and $R_n^B$ for different wind generation levels is also shown. This is useful when the value of risk, for a specific line or bus outage, needs to be observed. Moreover, it can aid in studying the impact of the wind generation level on the risk value for a specific line or bus contingency. Various existing research [30-41] indicates the significance of analyzing the impact of wind generation on power system security assessment, in the presence of uncertainty.

It must be mentioned that there are a few challenges and limitations for the practical implementation of the proposed risk-based security assessment approach. Firstly, modeling of the uncertainty in the associated random variables (load, wind generation, etc.) is a complicated process and selecting an optimal method in this regard is a challenging task. The uncertainty in these variables can greatly impact the final value of risk. Secondly, there are no established standard risk metrics which the utilities can use. This creates a sense of confusion as there is no uniformity in assessing the consequence (impact) of an undesired event. Also, there are very limited softwares which can be used to simulate and analyze the power system probabilistic processes for evaluation of risk.

## VIII. Conclusion and Future Work

In this paper, a novel security indicator based on average cost deviation from the mean optimal system operating cost was proposed for a standard test power transmission system (IEEE 39-bus). The impact of wind generation levels was observed on the security of the transmission test system, using both linear and quadratic wind cost functions. It was found that higher (100%) wind penetration is beneficial in improving system security, as it reduces risk and average cost deviation, considering *(N-1), (N-2), (N-3)* line outages and *(N-1)* bus outages. In contrast, wind penetration of 52% deteriorates the system security as it increases values of risks and average cost deviations. Moreover, the results obtained helped rank the most and least critical line(s) and most and least critical buses in the system for different levels of wind generation. The results obtained are of great significance to power system planners as they help to ascertain the impact of wind generation penetration levels on the security of lines and buses, in terms of risk. Thus, this analysis can help planners make efficient decisions.

As a future work, the proposed framework, being generic and universal, can easily be extended to a large-scale power system, where the system can be divided into multiple regions to evaluate the enclosed risk index in each region to establish a global index. The SCOPF problem becomes quite computationally challenging for a large-scale system with numerous uncertainties. In this regard, machine learning and deep learning approaches for SCOPF are still an open area of research whose full potential is yet to unravel. Further work may also include research on algorithms to expand optimal corrective and preventive control actions for reducing risk.

**Disclosure statement**

No potential conflict of interest was reported by the author.

**Data availability statement**

Data available on request from the author.

**Funding statement**

No funding was received.

**Notes on Contributor**

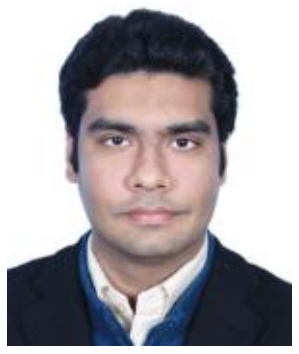

**Umair Shahzad** is currently working as a Lecturer at the School of Computer Science and Engineering, University of Sunderland, UK. In 2021, he received the Ph.D. degree in Electrical Engineering from The University of Nebraska-Lincoln, USA. Moreover, he received a B.Sc. Electrical Engineering degree from the University of Engineering and Technology, Lahore, Pakistan, and a M.Sc. Electrical Engineering degree from The University of Nottingham, England, in 2010 and 2012, respectively. His research interests include power system security assessment, power system stability, machine learning, and probabilistic methods applied to power systems.